\documentstyle [12pt]{article}
\newcommand{\f}{\frac}
\newcommand{\Om}{\Omega}

\newcommand{\bfx}{{\bf x}}
\newcommand{\bfv}{{\bf v}}
\newcommand{\bfr}{{\bf r}}
\newcommand{\bfk}{{\bf k}}
\newcommand{\bc}{\begin{center}}
\newcommand{\ec}{\end{center}}
\newcommand{\be}{\begin{equation}}
\newcommand{\ee}{\end{equation}}
\newcommand{\lan}{\langle}
\newcommand{\ran}{\rangle}
\newcommand{\etal}{{\it et al.~}}

\title{{\huge {\bf Velocity Differences as a Probe}}\\
  {\huge {\bf of Non--Gaussian Density Fields}} }
\author{ {\bf Paolo CATELAN}$^{1,2}$ and {\bf Robert J. SCHERRER}$^{3}$\\ \\
$^{1}$ {\it SISSA -- International School for Advanced Studies,} \\
{\it Strada Costiera 11, 34014 Trieste, Italy} \\ \\
$^{2}${\it Department of Physics, Astrophysics} \\
{\it Keble Road, Oxford OX1 3RH, UK ~[after March 1, 1994]} \\ \\
$^{3}$ {\it Department of Physics, The Ohio State University} \\
{\it Columbus, OH 43210} }
\date{}
\begin{document}
\maketitle
\bc
{\it Astrophysical Journal}, submitted
\ec
\bc
{\it SISSA Ref. 37/94/A}
\ec
\bc
{\it OSU--TA--4/94}
\ec

\newpage
\vspace{1cm}

\section*{\center Summary}
We examine the multi--point velocity field for non--Gaussian models as a probe
of non--Gaussian behavior.  The two--point velocity correlation is not
a useful indicator of a non--Gaussian density field, since it depends only on
the power spectrum,
even for non--Gaussian models.  However, we show that the distribution
of velocity differences $\bfv_1 - \bfv_2$, where $\bfv_1$ and
$\bfv_2$ are measured at the points $\bfr_1$ and $\bfr_2$, respectively, is a
good probe of non--Gaussian
behavior, in that $p(\bfv_1 - \bfv_2)$ tends to be non--Gaussian whenever the
density
field is non--Gaussian.  As an example, we examine the behavior of $p(\bfv_1 -
\bfv_2)$
for non--Gaussian seed models, in which the density field is the convolution of
a distribution of points with a set of density profiles.  We apply these
results
to the global texture model.

\vspace{1cm}
\noindent{\bf Key Words:} Cosmology -- dark matter -- galaxies: clustering
-- galaxies -- large--scale Structure of the Universe.

\newpage
\section{Introduction}
A fundamental problem in modern cosmology is to understand how the
large scale structures in the observed Universe arose from the
primordial energy density fluctuations. The simplest
hypothesis is that an inflationary de Sitter phase produced
Gaussian, scale--invariant adiabatic fluctuations leading to the
generation of large scale clustered structure grew due to their
gravitational instability. This scenario, in a Universe
dominated by cold dark matter (CDM), has been considered the most
succesful theory for the formation of galaxy or cluster structures in
the Universe (Frenk {\em et al.} 1988; Davis {\em et al.} 1992).

However, there are also notable setbacks for this model from observational
evidence.
The most serious of these observational challenges have come
from the variance in the cell counts of $IRAS$ galaxies (Efstathiou
{\it et al.} 1990), from the measurement of the skewness in the distribution of
QDOT--$IRAS$ galaxies out to 140$\,h^{-1}$Mpc (Saunders {\it et al.} 1991),
from the large--scale clustering of radio galaxies (Peacock \& Nicholson 1991)
and rich clusters of galaxies (Efstathiou {\it et al.} 1992), and from the
statistical analysis of the APM galaxy distribution on scales beyond
10$\,h^{-1}$Mpc (Maddox {\it et al.} 1990).

Occurrence of high--redshift quasars
(Warren {\it et al.} 1987), radio
sources (Lilly 1988; Chambers, Miley
\& van Breugel 1990) and protoclusters
of galaxies (Uson, Bagri \& Cornwell
1991), as well as the Great Bootes
Void, Great Wall, Great and Giant
Attractor regions (Kirshner {\it
et al.} 1981; Geller \& Huchra 1990;
Lynden--Bell {\it et al.} 1988;
Scaramella {\it et al.} 1989) may
pose further problems to the ``standard''
CDM Gaussian model (see the discussion
in Davis {\it et al.} 1992).

As a consequence, alternative theories of the origin of large scale structure
in the Universe continue to be worth investigation. Among the
chief rivals to the inflationary model are a class of models in which
the initial inhomogeneities are
produced by a primordial population of {\it seeds}, e.g. cosmic string loops
(Zel'dovich 1980; Vilenkin 1981), global textures (Gooding et al. 1992 and
references
therein), or generic ``seeds" (Villumsen et al. 1991; Gratsias et al. 1993),
and the
final linear matter fluctuations are not described by a Gaussian
distribution. Non--Gaussian distributed perturbations are the
most general statistical framework for computing cosmological observables, such
as spatial galaxy correlation functions (Matarrese, Lucchin \& Bonometto 1986;
Scherrer \& Bertschinger 1991), expected size and frequency of high density
regions (Catelan, Lucchin \& Matarrese 1988), hotspots and coldspots
in the cosmic microwave background (CMB) distribution on the sky
(Coles \& Barrow 1987; Kung 1993) and
higher order temperature correlation functions (Luo \& Schramm 1993;
Gangui {\it et al.} 1994).
The importance of the signature of the initial distribution
of non--Gaussian density fluctuations for the final clustering pattern of the
cosmic structures has been recently emphasized, in a series of N--body
simulations, by Messina {\it et al.} (1990), Moscardini {\it et al.} (1991),
Matarrese {\it et al.} (1991) and Weinberg \& Cole (1992).
Coles {\it et al.} (1993) show that the mass distribution in models with
initially non--Gaussian fluctuations exhibits systematic departures from
Gaussian behavior on intermediate to large scales.

A powerful method to probe the matter distribution for non--Gaussian
behavior
is to examine the large--scale bulk velocity field, which is, in the linear
regime
(i.e. large scales), directly related to the density field (Peebles 1980). The
problem has been addressed by Scherrer (1992).  Scherrer, noting that the
RMS velocities are the same function of the power spectrum for both
Gaussian and non--Gaussian density fields, examined the distribution of
velocities for non--Gaussian density fields.  For density fields
in which the gravitational potential field is non--Gaussian by construction
(``local" non--Gaussian fields) the velocity field can be strongly
non--Gaussian.
However, Scherrer found that the one--point distribution of linear velocities
for
``seed" models, in which the density field is the convolution of
a discrete set of points with a set of density profiles, is highly
Gaussian even when the density field is strongly non--Gaussian;
this result has also been seen in numerical simulations of the
texture model (Gooding, et al. 1992).  The reason for this is the
central limit theorem:  the linear velocity field is an integral over the
density field; if the density field is sufficiently uncorrelated,
the sum of uncorrelated densities produces a Gaussian velocity field.
Of course, the linear velocity field in this case cannot be an exact
multivariate
Gaussian, as the divergence of such a field (which is proportional to the
density
field) would then have to be exactly a Gaussian.

In this paper, we exploit this fact to derive a velocity statistic which
is sensitive to non--Gaussian density fields.  We extend the
analysis of Scherrer (1992) to derive the two--point velocity
distribution for seed models.  We show that the distribution of
velocity {\it differences}
is an effective probe of non--Gaussian
behavior in the density field.

In the next section, we review the general properties of linear cosmological
velocity fields, valid for any mass density
distribution, Gaussian or non--Gaussian.
In  $\S$3, after reviewing the statistical properties
of seed models, we work out exact analytical expressions for
the 1--point and 2--point velocity distributions using the partition function
for randomly distributed seed--objects.  These results are used to derive an
expression
for the distribution of the difference between velocities measured at two
different points.
We apply our results first
to a toy seed model to illustrate some of our general arguments about the
usefulness
of this method, and then to the
global texture model with cold
dark matter (Gooding {\it et al.} 1992).  Our results
are discussed in Section 4.

\section{Cosmological Velocity Fields}
During the linear regime, when the amplitude of matter fluctuations is
very small ($\delta \ll 1$), the peculiar velocity ${\bf v}$ is related
to the matter fluctuation $\delta$ by the equation (see, e.g., Peebles 1980)
\be
\nabla\cdot{\bf v}({\bf r},t)=- H\,\Om^{0.6}a\;\delta({\bf r}, t)\;,
\ee
where $H$ is the Hubble parameter, $a$ is the scale factor, and $\Om$ is
the mean density in units of the critical density.
Any transverse mode, defined by the condition $\nabla\cdot{\bf v}=0$ and
corresponding to the rotational part of the velocity field, decays with the
expansion (namely, the vorticity decays with rate $\propto a^{-2}$ in an
Einstein--de Sitter universe). Therefore, the integral expression
of the solution of eq. (1) may be written as (Peebles 1980)
\be
{\bf v}({\bf r},t) = \f{H\,\Om^{0.6}a}{4\pi}\;\nabla\int d^3{\bf r}'
\f{\delta({\bf r}',t)}{|{\bf r}-{\bf r}'|}=
-\f{2}{3}\f{\Om^{-0.4}}{Ha}\,\nabla\phi({\bf r}, t)\,,
\ee
which is manifestly irrotational; $\,\phi\,$ is the peculiar gravitational
potential. We see that in the linear regime, both density and velocity fields
may be derived from a unique potential $\phi\,$.
In eqs. (1)--(2), we consider only the growing modes
for density and velocity.  In what follows, we
fix $t$ equal to the present time, taking
$a=1$ and $H = H_0$.

Results for the linear velocity field are of interest because the observations,
when smoothed on sufficiently large scales, recover this linear velocity
field.
This smoothing is
usually done by filtering on the scale $R$ the velocity field
${\bf v}({\bf r})$ by means of a {\it window function} $W_R(x)$,
$x\equiv|{\bf x}|$,
\be
{\bf v}_R({\bf r})=\int\!d^3{\bf x}\,{\bf v}({\bf x})\,W_R(|{\bf x}-{\bf r}|).
\ee
The windowing convolution averages the velocity field ${\bf v}$ over the
point ${\bf x}$ in a volume $\sim R^{\,3}$, where one has no interest in the
substructure. Typical choices are the {\it Gaussian} and the {\it top--hat}
window functions (e.g. Bardeen {\it et al.} 1986).
Working with the Fourier transforms of the quantities in eq. (3), it
is easy to show that the smoothed linear velocity field corresponding to
a particular density field is identical to the unsmoothed linear velocity
field corresponding to that density field smoothed with the same
window function (Scherrer 1992). Thus, the smoothed
velocity field can be derived by applying eq. (2) to the smoothed density
field.

An equivalent manner to describe the velocity field is
in terms of the Fourier components of the density field, $\widehat{\delta}({\bf
k})$. An important
cosmological observable, the rms peculiar velocity $\sigma_{{\bf v}}$, can be
expressed very simply in terms of the power spectrum $P(k)$:
\be
\sigma_{\bf v}^2\equiv\lan{\bf v}\cdot{\bf v}\ran=
H_0^2\,\Om^{1.2}\int_0^{\infty}4\pi ~P(k) ~dk,
\ee
where $P(k)$ is defined by
$\lan\widehat{\delta}({\bf k}_1)\widehat{\delta}({\bf k}_2)\ran\equiv
\delta_D({\bf k}_1+{\bf k}_2)\,P(k_1)$.
The relation between the rms velocity
$\sigma_{{\bf v}}$ and the power spectrum $P(k)$ is valid for any type of
density field, Gaussian or non--Gaussian (Scherrer 1992).  Thus $\sigma_{\bf
v}$
cannot be used to distinguish Gaussian from non--Gaussian models.  For this
reason Scherrer (1992) was motivated to examine the {\it distribution} of
${\bf v}$, which is different for Gaussian and non--Gaussian models.
For example, in local non--Gaussian models, in which the potential field
is local nonlinear function of a Gaussian field (Kofman et al. 1989; Moscardini
et al. 1991)
the one--point distribution of a single component of $\bf v$ is strongly
non--Gaussian (Scherrer 1992).  However, for seed
models, in which the density field is the convolution of a density profile with
a random
distribution of points, the distribution of a single component of $\bf v$ is
nearly Gaussian.

We may look for better tracers of a non--Gaussian density field among
the multi--point velocity distribution.  An obvious choice is the two--point
velocity correlation function
\be
\langle \bfv(\bfx) \cdot \bfv(\bfx+\bfr) \rangle = H_0^2 \Omega^{1.2} \int
{1\over k^2} P(k) e^{i \bfk \cdot \bfr} d^3 \bfk.
\ee
We see that just as in the case of the rms velocity, the two--point velocity
correlation depends only on the power spectrum.  Thus, it is incapable of
distinguishing a Gaussian density field from a non--Gaussian field having
the same power spectrum.

We argue here that a better test for non--Gaussian behavior is the distribution
of velocity differences $p(\bfv_1 - \bfv_2)$, where $\bfv_1$ and $\bfv_2$ are
the
velocities measured at the points $\bfr_1$ and $\bfr_2$, respectively.
In the usual way,
we define $v_\parallel$ and $v_\perp$ to be the components of the velocity
parallel
and perpendicular to $\bfr_1-\bfr_2$, respectively.  Then it is plausible that
the distribution of $v_{1\parallel} - v_{2\parallel}$
will become non--Gaussian for a non--Gaussian density field in the limit where
$|\bfr_1 - \bfr_2| \rightarrow 0$.
[We measure $v_{1\parallel} - v_{2\parallel}$ relative to a coordinate system
in which the direction of increasing distance points from $\bfr_2$ to $\bfr_1$;
this insures that $v_{1\parallel} - v_{2\parallel}$ has the same sign as
$\partial v_x/\partial x$ in the limit $|\bfr_1 - \bfr_2| \rightarrow 0$.]
{}From eq. (1), we know that
the distributions of
$\nabla \cdot \bfv$ and $\delta$ must be the same (up to a multiplicative
constant).
For an isotropic density field, $\partial v_x/\partial x$, $\partial
v_y/\partial y$,
and $\partial v_z/\partial z$ must all have the same distribution.  There is no
simple
relationship between the distribution of
$\partial v_x/\partial x$ and
the distribution of $\delta$, since the various partial derivatives of $\bfv$
are correlated.  However, it would require a very contrived density field
for $\partial v_x/\partial x$, $\partial v_y/\partial y$ and
$\partial v_z/\partial z$ to all have a nearly Gaussian one--point
distribution,
while $\delta$ was highly non--Gaussian.  Furthermore,
$v_{1\parallel} - v_{2\parallel}$ is proportional to
$\partial v_x/\partial x$ in the limit where the separation between $\bfr_1$
and $\bfr_2$ goes to zero.  Thus, we expect $p(v_{1\parallel} -
v_{2\parallel})$ to
be non--Gaussian for sufficiently small separations
whenever $p(\delta)$ is non--Gaussian.  This is our motivation for
examining $p(\bfv_1 - \bfv_2)$ as a probe of non--Gaussian behavior.  In
essence,
the taking of velocity differences is a poor man's derivative.  Integrating
over
the density field to derive the velocity field can drive $p(\bfv)$ to a
Gaussian
because of the central limit theorem; taking velocity difference restores the
non--Gaussian behavior lost through integration.  Of course, a major issue is
the
maximum separation over which non--Gaussian behavior can be detected in the
distribution of $\bfv_1 - \bfv_2$; if $p(\bfv)$ is Gaussian, and if velocity
correlations
go to zero at sufficiently large distances, then the $p(\bfv_1 - \bfv_2)$
must also go to a Gaussian over those distances.

\section{Velocity Differences in Seed Models}

To test our velocity--difference statistic,
we analyze the large--scale velocity field in a special class of
intrinsically non--Gaussian models, in which the initial linear
density field can be described as the convolution of a set of density profiles
with
a random distribution of points (Scherrer \& Bertschinger 1991; Scherrer 1992).
This set of non--Gaussian models provides a useful test of our method
for several reasons.  Such distributions are analytically tractable and can
describe, for example, the density field produced by textures (Gooding et al.
1992) and
the seeded hot dark matter model (Villumsen et al. 1991; Gratsias et al. 1993).
Further, such models tend to produce a Gaussian one--point velocity
distribution
even when the density field is strongly non--Gaussian (Scherrer 1992), so they
are useful
to determine if the velocity differences provide a better probe of the
non--Gaussian behavior.

In these seed models, the linear density field
$\delta(\bfr)$ may be written as a convolution of a given density profile
$f(m,\bfr)$
with a set of points having mean density $n(m)$, where $m$ is simply
a parameter or set of parameters such as mass or formation time:
\be
\delta(\bfr)=\sum_h f(m_h,\bfr - \bfx_h)
\ee
where the sum is taken over all seeds with positions $\bfx_h$.
The statistical properties of the density field are entirely determined
by $f(m,\bfr)$, $n(m)$, and the spatial correlations of the seeds
(see Scherrer \& Bertschinger 1991 for more details).  Here we
follow Scherrer (1992) and consider only randomly distributed seeds for
which $f$ is spherically symmetric: $f(m,\bfr) = f(m,r)$.
This simpler model is a reasonably
good description of the texture model (Gooding et al. 1992) and the
seeded hot dark matter model (Villumsen et al. 1991; Gratsias et al. 1993).

In the linear regime, both the density perturbations and the velocities
induced by the seeds can be added linearly, so that
the perturbation $\delta({\bf r})$ in eq. (1) produces at ${\bf r}$
the peculiar velocity ${\bf v}({\bf r})$ given by
\be
{\bf v}({\bf r}) = \sum_h\widetilde\bfv(m_h,\bfr - \bfx_h),
\ee
where $\widetilde\bfv(m,\bfr-\bfx)$ is the contribution to ${\bf v}$,
measured at the point ${\bf r}$, due to a single $m$--seed located at the
point ${\bf x}$.  From eq. (1), we have
\begin{equation}
\nabla\cdot\widetilde\bfv(m,\bfr-\bfx) = -H_0\,\Om^{0.6}\;
f(m,|{\bf r}-{\bf x}|)\;;
\end{equation}
Thus, $f(m,|{\bf r}-{\bf x}|)$ acts exactly as a density
perturbation source for the single seed--velocity contribution
$\widetilde\bfv(m,\bfr-\bfx)$, and the total velocity ${\bf v}({\bf r})$
is the superposition, for the entire population of seeds, of the single
seed velocities.
To determine $\widetilde\bfv(m,\bfr-\bfx)$, we integrate the previous
equation on the sphere $V_{\bf x}$ centered in ${\bf x}$. Gauss's law
gives
\be
\widetilde\bfv(m,\bfr-\bfx)=
-\beta\,\bar{f}(m,|{\bf r}-{\bf x}|)\,
\f{{\bf r}-{\bf x}} {|{\bf r}-{\bf x}|}\;,
\ee
where $\beta\equiv H_0\,\Om^{0.6}$:
we see that the $m$--seed at ${\bf x}$ produces a velocity component
$\widetilde \bfv(m,\bfr-\bfx)$ at ${\bf r}$ which points {\it toward}
the seed location, i.e., each seed tends to create a spherical velocity
field around it. In eq. (9), we have defined the quantity
\be
\bar{f}(m,|{\bf r}-{\bf x}|)\;\equiv \;\f{1} {|{\bf r}-{\bf x}|^2}\,
\int_0^{|{\bf r}-{\bf x}|}\!\! dr' \, r'^2 \, f(m,r')\;,
\ee
which represents the integrated density perturbation due to the seed
at ${\bf x}$, enclosed in a sphere of radius $|{\bf r}-{\bf x}|$ centered
on the seed.
{}From Eqs.(7) and (9) we obtain the total velocity $\bfv$ in term of the
accretion pattern,
\be
{\bf v}({\bf r}) = -\beta \sum_h \bar{f}(m_h,|{\bf r}-{\bfx_h}|)\,\f{{\bf
r}-{\bfx_h}} {|{\bf r}-{\bfx_h}|}.
\ee
Equations (7)--(11), are somewhat more general
than those given by Scherrer (1992), in that the location of the seed is
arbitrarily
chosen.

We now analyze the statistical properties of the velocity field $\bfv$ for
these seed
models.
If the seeds are randomly distributed, the probability
distribution of the total velocity $\bfv$ may be obtained
by applying the Poisson model (Peebles
1980; Fry 1985).  To do this, we introduce the stochastic Poisson
variables $\epsilon_h$ in such a way that
\be
\epsilon_h =
\left\{
\begin{array}{ll}
1 & \mbox{if a seed $\in d{\bf x}_h$} \\
0 & \mbox{otherwise}\;,
\end{array}
\right.
\ee
with mean
\be
\lan\epsilon_h\ran = n(m_{\,h})\,dm_{\,h}\,d{\bf x}_{\,h}\;.
\ee
{}From eq.(7), we know that ${\bf v}({\bf r})$ is the superposition of all the
single seed contributions,
and to remember the stochastic information, i.e., presence or absence
of a seed in $d{\bf x}_h$, we define the variable
$\widehat{{\bf v}}({\bf r})\equiv\sum_h\widetilde{{\bf v}}(m_h,\bfr-{\bf
x}_h)\epsilon_h$.

The probability distribution function $p({\bf v})$ may be calculated according
to the definition (see, e.g., Ma 1985)
\be
p({\bf v})=
\lan\,\delta_D[\widehat{{\bf v}}({\bf r})-{\bf v}({\bf r}) ]\,\ran
=\f{1}{(2\pi)^3}\int_{-\infty}^{+\infty}d^3{\bf {\bf t}} \;
{\rm e}^{-i\,{\bf {\bf t}}\cdot{\bf v}({\bf r})}\,\prod_h
\phi_{\epsilon_h}({\bf t})\;,
\ee
where $\phi_{\epsilon_h}({\bf t})$ is the characteristic (or moment generating)
function for the {\it single} seed discrete process (Peebles 1980; Fry 1985)
\be
\phi_{\epsilon_h}({\bf t})\equiv
\lan{\rm e}^{-\,i{\bf t}\cdot
\widetilde{\bf v}(m_h,{\bf r}-\bfx_h)\epsilon_h} \ran
= {\rm exp}\; \left[n(m_{\,h})\,dm_{\,h}\,d^3{\bf x}_{\,h}\left(
{\rm e}^{\,i {\bf {\bf t}} \cdot \widetilde{\bf v}(m_h,\bfr-\bfx_h)} -
1\right)\right]
\ee
Therefore, the characteristic function for the total velocity is the product of
the
individual $\phi_{\epsilon_h}$
\be
\phi({\bf t})=\prod_h\phi_{\epsilon_h}({\bf t})
\longrightarrow
{\rm exp}\,\int n(m)\,dm\,d^3{\bf x}\left(
{\rm e}^{\,i {\bf t} \cdot \widetilde\bfv(m,\bfr-\bfx) } - 1\right)
\ee
in the continuum limit $dm_h\rightarrow0$ and  $d{\bf x}_h\rightarrow0$.
Finally, the velocity pdf $p({\bf v})$ may be written as
\be
p({\bf v})=\f{1}{(2\pi)^3}\int_{-\infty}^{+\infty}d^3{\bf {\bf t}} \,
{\rm e}^{-i\,{\bf {\bf t}}\cdot{\bf v}}
\,{\rm exp}\,\int n(m)\,dm\,d^3{\bf x}\left(
{\rm e}^{\,i {\bf t} \cdot \widetilde\bfv(m,\bfr-\bfx) } - 1\right)\;,
\ee
which is a generalization of the result given by Scherrer (1992).

Scherrer (1992) showed that the
distribution of the one point velocity, $p(\bfv)$,
is a rather poor indicator of non--Gaussian behavior for seed models
with randomly distributed seeds. In particular, for the
seeded hot dark matter model,
a non--Gaussian velocity field requires an extremely low seed density,
while the texture model with cold dark matter has a velocity field which is
nearly indistinguishable from a Gaussian.

We now consider the two--point distribution of velocities,
$p(\bfv_1, \bfv_2)$.
Taking advantage of our
previous results, it is straightforward to show that the
the 2--point distribution is given by
$$
p({\bf v}_1, {\bf v}_2)=
\lan\,
\delta_D[\widehat{{\bf v}}({\bf r}_1)-{\bf v}({\bf r}_1) ]
\,\delta_D[\widehat{{\bf v}}({\bf r}_2)-{\bf v}({\bf r}_2) ]
\,\ran
$$
\be
=\f{1}{(2\pi)^6}\int_{-\infty}^{+\infty}d^3{\bf {\bf t}}_1\,
d^3{\bf {\bf t}}_2 \;
{\rm e}^{ -i\,{\bf {\bf t}}_1\cdot{\bfv_1}
-i\,{\bf {\bf t}}_2\cdot{\bfv_2}}\;
\phi({\bf t}_1, {\bf t}_2)\;,
\ee
where
\be
\phi({\bf t}_1, {\bf t}_2)\equiv
{\rm exp}\,\left[\int n(m)\,dm\,d^3{\bf x}\left(
{\rm e}^{ \,i {\bf {\bf t}}_1 \cdot \widetilde{\bf v}(m,\bfr_1-\bfx) +
i {\bf {\bf t}}_2 \cdot \widetilde{\bf v}(m,\bfr_2-\bfx)} - 1\right)\right]\;,
\ee
is the characteristic function for the 2--point distribution
of velocities;
it generates the 2--point velocity correlation tensor component
defined by (Groth, Juszkiewicz \& Ostriker 1989)
$\xi_{\alpha\beta}(\bfr_1,\bfr_2)\equiv\lan\,v_{\alpha}({\bfr_1})\,
v_{\beta}({\bfr_2})\,\ran\;$,
$\;\alpha,\;\beta = 1,\; 2,\; 3\;$,
through partial differentiation:
\be
\xi_{\alpha\beta}(\bfr_1,\bfr_2) = - {\partial \phi \over \partial
t_{1\alpha}}(0)
{\partial \phi \over \partial t_{2\beta}}(0).
\ee
The statistical properties of the large--scale peculiar velocity field, as
described by the correlation tensor, are extensively examined in
Gorsky \etal (1989) and Landy \& Szalay (1992).

{}From eq. (18), one can in particular obtain the probability
distribution of the {\it velocity difference}
\be
p({\bf v}_1-{\bf v}_2)
=\f{1}{(2\pi)^3}\int_{-\infty}^{+\infty}d^3{\bf {\bf t}}\;
{\rm e}^{ -i\,{\bf {\bf t}}\cdot({\bfv_1}-{\bfv_2})}\;
\phi({\bf t})
\ee
where now
\be
\phi({\bf t})=
{\rm exp}\,\left[\int n(m)\,dm\,d^3{\bf x}\left(
\,{\rm e}^{ \,i {\bf {\bf t}}\cdot [\widetilde{\bfv}(m,\bfr_1-\bfx) -
\widetilde{\bfv}(m,\bfr_2-\bfx)]} - 1\right)\right]\;.
\ee
A convenient measure of the deviation from a Gaussian is given by the cumulants
$\kappa_p$, defined by
\be
\ln \phi(t) = \sum_{p=1}^\infty \kappa_p {(it)^p \over p!}
\ee
For the distribution of velocity differences, it is clear
that
\be
\kappa_p = \int n(m)~ dm ~d^3 \bfx ~[\widetilde{v}_\alpha(m,\bfr_1-\bfx) -
\widetilde{v}_\alpha(m,\bfr_2-\bfx)]^p
\ee
where eq. (24) gives the cumulants corresponding to the distribution
of the $\alpha$ component of the velocity differences.  The normalized
cumulants
\be
\lambda_p \equiv \kappa_p /\kappa_2^{p/2}
\ee
give a measure of the deviation of a distribution from a Gaussian; for
$\lambda_p \ll 1$ the distribution is nearly Gaussian, while $\lambda_p \gg 1$
indicates a highly non--Gaussian distribution.

We argued in $\S$2 that the distribution of $\bfv_1-\bfv_2$ should show the
same
level of non--Gaussian behavior as the distribution of $\delta$; here we test
these claims numerically.  First consider a toy seed model which consists
of randomly--distributed seeds with number density $n_0$, all having the same
mass $m_s$, so that $n(m) = n_0 \delta_D(m-m_s)$.  We also assume
a spherical tophat density profile:
$f(m,r) = mg_0$ for $r \le r_0$, and $f(r) = 0$ for $r > r_0$.  [This is not
a physically realistic model, but the distributions for $\delta$ and $\bf v$
can be derived analytically, allowing us to confirm some of our general
arguments
about the usefulness of the velocity difference statistic.
Some of the properties of this model
were examined by Scherrer \& Bertschinger (1991).  Alternately, this model
gives the velocity
field smoothed with a spherical tophat window function
for a set of point--like seed masses].
To illustrate the points discussed
earlier, we calculate $\lambda_p$ for the distributions of $\delta$, $v_x$ (a
single component of the velocity)
and $\partial v_x/\partial x$.  If we define $\bar n = n_0 4\pi r_0^3/3$, then
the distribution
of $\delta$ is just a Poisson distribution with
\be
\lambda_p = \bar n^{1-p/2}
\ee
The value of $\lambda_p$ for $v_x$ can be derived using the techniques outlined
in Scherrer (1992); we find (for even $p$)
\be
\lambda_p = \bar n^{1-p/2}\Biggl({3 \over p+1}\Biggr)\Biggl({1\over
p+3}+{1\over 2p-3}
\Biggr)\Biggl({5\over 6}\Biggr)^{p/2}.
\ee
Finally, consider the distribution of $\partial v_x/\partial x$, which is
proportional to
$v_{1\parallel} - v_{2\parallel}$ when the separation goes to 0.
In this case, eq. (24) with $\widetilde{v}_\alpha(m,\bfr_1-\bfx) -
\widetilde{v}_\alpha(m,\bfr_2-\bfx)$ replaced by $\partial\widetilde
v_x/\partial x$
gives
\be
\lambda_p = \bar n^{1-p/2} (5/9)^{p/2}\biggl[(-1)^p+{1 \over 2p-2} \int_{-1}^1
(3x^2-1)^p dx\biggr].
\ee
The lowest order non--zero deviation from a Gaussian for $v_x$ is given by
$p=4$.
For this case we find that $\lambda_4 = \bar n^{-1}$ (distribution of
$\delta$),
$\lambda_4 = 0.14 \bar n^{-1}$ (distribution of $v_x$), and
$\lambda_4 = 0.45 \bar n^{-1}$ (distribution of $\partial v_x/\partial x$).
Furthermore,
in the limit where $p \rightarrow \infty$, the asymptotic
expressions for $\lambda_p$ are
\be
\lambda_p = \bar n^{1-p/2} {9 \over 2p^2}\Biggl({5\over 6}\Biggr)^{p/2},
\ee
for $v_x$, and
\be
\lambda_p = \bar n^{1-p/2}{1\over 3p^2} \Biggl({20 \over 9}\Biggr)^{p/2},
\ee
for $\partial v_x/\partial x$.
Thus,
for an appropriate choice of $\bar n$, the distribution of $v_x$ can
be nearly Gaussian while the distribution of $\delta$ is highly
non--Gaussian, in agreement with the results of Scherrer (1992).  However,
the distribution of $\partial v_x/\partial x$ displays roughly the same
deviation from
Gaussianity as does the distribution of $\delta$.  Therefore, for sufficiently
small
separations, the distribution of $v_{1\parallel}-v_{2\parallel}$ will also
trace the non--Gaussian behavior of $\delta$.

Now consider a physically realistic non--Gaussian model:
the $\Omega_0 = 1$ CDM
global texture model (Gooding et al. 1992 and references therein ).
This model can be approximated as a seed model, with a random distribution
of seeds (the locations of the texture knots).  We take the parameter
$m$ to be the conformal time $\tau$ at which the knots unwind; then
the number density of knots unwinding at $\tau$ is
(Gooding et al. 1992)
\be
dn = \nu \tau^{-4}d \tau,
\ee
with $\nu = 0.04$.
We use an analytic approximation for $f(\tau,r)$ given by Gooding (1992)
\be
f(\tau,r) = C {e^{-\lambda r/\tau}(1-\lambda r/2\tau) \over r}f_g,
\ee
where $\lambda = 1.88$ and the constant $C$ does not affect our calculations.
The growth factor $f_g$ is
\be
f_g = a_i[\delta_2(a_i)\delta_1(a) - \delta_1(a_i)\delta_2(a)],
\ee
where $a$ is the scale factor at which the density perturbations are measured,
$a_i$ is the scale factor corresponding to the conformal time $\tau$,
and $\delta_1$ and $\delta_2$ are the growing and decaying modes in linear
theory (see Scherrer 1992 for more details).

Using this approximation for the CDM texture model, we have numerically
evaluated
the normalized skewness $\lambda_3$
of the distribution of the component of the velocity difference
along the separation $\bfr_1-\bfr_2$,
$v_{\parallel}(\bfr_1)-v_{\parallel}(\bfr_2)\,$, as a function of
separation distance $|\bfr_1 - \bfr_2|$.
In Figure 1, we plot this skewness
versus
the separation distance of the points at which the velocities are measured
(solid curve).
We see that for relatively small separations, the difference
of velocities {\it is} a good tracer of non--Gaussian behavior. Our
statistic is useful up to a separation of about 10$h^{-2}$ Mpc.  The density
field
for the texture model has positive skewness, so it is not surprising
that that $v_\parallel(\bfr_1) - v_\parallel(\bfr_2)$ has negative
skewness (see eq. 1 and our argument in $\S$2).
It is trivial to extend these results to the case of the smoothed velocity
field because,
as noted in \S2, the smoothed linear velocity field is identical to the
unsmoothed
linear velocity field corresponding to the smoothed density field.  For a
velocity field
smoothed with a
Gaussian window function $W(r) = \exp(-r^2/r_0^2)$, the
skewness of the
distribution of velocity differences along the separation vector is given in
Figure 1 for $r_0 = 2h^{-2}$ Mpc (dashed curve).
We find, surprisingly, that the smoothed velocity field actually deviates more
strongly
from a Gaussian than the unsmoothed field.
The density fluctuation produced by a single texture integrated out to a radius
$R$ goes to zero in the limit $R \gg \tau$
(Gooding, et al. 1992) so smoothing a texture with a window function of radius
$r_0 \gg \tau$
effectively erases the texture, and we are left with a density
field consisting of larger textures with larger separations, giving a more
non--Gaussian
field.  Thus, the enhancement of
the non--Gaussian nature of $p(v_\parallel(\bfr_1) - v_\parallel(\bfr_2))$
with smoothing
is probably peculiar to the texture
model.

\section{Discussion}

Our results indicate that the distribution of $v_\parallel(\bfr_1) -
v_\parallel(\bfr_2)$,
the component of the velocity difference at two points along the vector
separating the
observation points, is a useful probe of non--Gaussian behavior.  This
distribution
tends to be non--Gaussian whenever the underlying density field is
non--Gaussian,
in contrast to the one--point distribution of velocities, which tends to
be more Gaussian than the underlying density field (Scherrer 1992).  Although
we have examined only a single class of non--Gaussian models, our results
should be
generally applicable to any non--Gaussian model, because of our argument in
$\S$2.
We note that a related argument for detecting signatures of non--Gaussian
density fields in the
cosmic microwave background
has been given by Moessner, Perivolaropoulos,
and Brandenberger (1994).   They argue that the temperature {\it gradient}
of the CMB should provide a better test to distinguish the cosmic string
model then the distribution of the temperature field itself.

\vskip 1 cm

P.C. is grateful to S. Matarrese for helpful discussions.  P.C. was supported
in part
by the Italian Ministero della Ricerca Scientifica e Tecnologica and
by the Fondazione Angelo Della Riccia.  R.J.S. was supported in part by
the DOE (DE--AC02--76ER01545).

\newpage

\parindent 0pt
\section *{References}

\begin{trivlist}

\item[] Bardeen, J.M., Bond, J.R., Kaiser, N., Szalay, A.S. 1986, ApJ,
304, 15

\item[] Catelan, P., Lucchin, F., \& Matarrese S. 1988,
Phys Rev Lett, 61, 267

\item[] Chambers, K.C., Miley, G.K., \& van Breugel, W. 1990, ApJ,
363, 21

\item[] Coles, P., \& Barrow, J.D. 1987, MNRAS, 228, 407

\item[] Coles, P., Moscardini, L., Lucchin, F., Matarrese, S., \& Messina,
A. 1992, MNRAS, 264, 749

\item[] Davis, M., Efstathiou, G., Frenk, C.S., \& White, S.D.M. 1992,
Nature, 356, 489

\item[] Efstathiou, G., et al., 1990, MNRAS,
247, 10P

\item[] Efstathiou, G., Dalton, G.B., Sutherland, W.J., \& Maddox S.J. 1992,
MNRAS, 257, 125

\item[] Frenk, C.S., White, S.D.M., Davis, M., \& Efstathiou, G., 1988,
ApJ,  327, 507

\item[] Fry, J.N. 1985, ApJ, 289, 10

\item[] Gangui, A., Lucchin, F., Matarrese, S., \& Mollerach, S. 1994,
ApJ, in press

\item[] Geller, M., \& Huchra, J. 1990, Science,  246, 897

\item[] Gooding, A.K., Park, C., Spergel, D.N., Turok, N., \& Gott, J.R. 1992,
ApJ, 393, 42

\item[] Gorski, K.M., et al. 1989, ApJ, 344, 1

\item[] Gratsias, J., Scherrer, R.J., Steigman, G. \& Villumsen, J.V. 1993,
ApJ,
405, 30

\item[] Groth, E.J., Juszkiewicz, R., \& Ostriker, J.P. 1989, ApJ,
 346, 558

\item[] Kirshner, R.P., Oemler, A., Schechter, P.L., \& Schetman, S.A. 1981,
ApJ,  248, L57

\item[] Kofman, L., Blumenthal, G.R., Hodges, H., \& Primack, J.R. 1989,
in Large--scale Structures and Peculiar Motions in the Universe, ed. D.W.
Latham
\& L.A.N. da Costa (San Francisco:  ASP), 339

\item[] Kung, J.H. 1993, Phys Rev D,  47, 409

\item[] Landy, S.D., \& Szalay, A.S., 1992, ApJ,  394, 25

\item[] Lilly, S.J. 1988, ApJ,  333, 161

\item[] Luo, X., \& Schramm, D.N. 1993, Phys Rev Lett, 71, 1124

\item[] Lynden--Bell, D., et al. 1988, ApJ, 326, 19

\item[] Ma, S.--K. 1985, Statistical Mechanics, World Scientific,
Singapore

\item[] Maddox, S.J., Sutherland, W.J., Efstathiou, G., \& Loveday, J. 1990,
MNRAS  242, 43P

\item[] Matarrese, S., Lucchin, F., \& Bonometto S. 1986, ApJ,
 310, L21

\item[] Matarrese, S., Lucchin, F., Messina, A., \& Moscardini L. 1991,
MNRAS,  253, 35

\item[] Messina, A., Moscardini, L., Lucchin, F., \& Matarrese, S. 1990,
MNRAS,  245, 244

\item[] Moessner, R., Perivolaropoulos, L., \& Brandenberger, R. 1994,
ApJ, 425, 365, 1994

\item[] Moscardini, L., Matarrese, S., Lucchin, F., \& Messina A. 1991,
MNRAS,  248, 424

\item[] Peacock, J.A., \& Nicholson, D. 1991, MNRAS,  253, 307

\item[] Peebles, P.J.E.  1980, The Large--Scale Structure of the Universe,
(Princeton:  Princeton University Press)

\item[] Saunders, W. et al. 1991, Nature,  349, 32

\item[] Scaramella, R., et al. 1989, Nature,  338, 562

\item[] Scherrer, R.J. 1992, ApJ,  390, 330

\item[] Scherrer, R.J., \& Bertschinger E. 1991, ApJ,  381, 349

\item[] Uson, J.M., Bagri, D.S., \& Cornwell, T.J. 1991, Phys Rev Lett,
 67, 3328

\item[] Vilenkin, A. 1981, Phys Rev Lett, 46, 1169, 1496(E)

\item[] Villumsen, J.V., Scherrer, R.J., \& Bertschinger E. 1991, ApJ,
367, 37; 381, 601(E)

\item[] Warren, S.J., Hewett, P.C., Osmer, P.S., \& Irwin, M.J. 1987,
Nature,  330, 453

\item[] Weinberg, D.H., \& Cole, S. 1992, MNRAS,  259, 652

\end{trivlist}

\newpage

\begin{center}
{\bf Figure Captions}
\end{center}

\vspace{1.5cm}
\noindent{\bf Figure 1.} The normalized skewness $\lambda_3$
of the distribution of $v_{\parallel}(\bfr_1)-v_{\parallel}(\bfr_2)$ [the
component of the velocity difference
parallel to the separation $\bfr_1-\bfr_2$],
as a function of the separation $|\bfr_1 - \bfr_2|$
for
the unsmoothed density field (solid curve) and the density field smoothed
with a Gaussian window function of radius $2h^{-2}$ Mpc (dashed curve).
\end{document}